\documentclass[twocolumn,prl,prb,nobibnotes,altaffilletter,amsmath,amssymb,amsfonts]{revtex4}
\date{\today}

\usepackage[latin1]{inputenc}
\usepackage{graphicx}
\usepackage{amsmath}
\usepackage{latexsym}
\usepackage{epsfig}

\begin{document}

\title{Generalized fluctuation-dissipation relation and effective temperature \\
       in off-equilibrium colloids}
\author{Claudio Maggi$^1$\footnote{Electronic address: cmaggi@ruc.dk}}
\author{Roberto Di Leonardo$^2$}
\author{Jeppe C. Dyre$^1$}
\author{Giancarlo Ruocco$^{3,2}$}
\affiliation{
$^1$DNRF Centre 'Glass and Time', IMFUFA, Dept. of Sciences, Roskilde University, DK-4000, Denmark. \\
$^2$Research center "Soft", INFM-CNR, c/o Dept. of Physics, "Sapienza" University of Rome, Italy. \\
$^3$Dept. of Physics, "Sapienza" University of Rome, Piazzale Aldo
Moro 5, I-00185, Rome, Italy. }

\begin{abstract}

The \emph{fluctuation-dissipation relation} (FDR), a fundamental
result of equilibrium statistical physics, ceases to be valid when
a system is taken out of the equilibrium. A generalization of FDR
has been theoretically proposed for out-of-equilibrium systems:
the kinetic temperature entering FDR is substituted by a
time-scale dependent effective temperature. We combine the
measurements of the correlation function of the rotational
dynamics of colloidal particles obtained via dynamic light
scattering with those of the birefringence response to study the
generalized FDR in an off-equilibrium clay suspension undergoing
aging. We i) find that FDR is strongly violated in the early stage
of the aging process and is gradually recovered as the aging time
increases and, ii), we determine the aging time evolution of the
effective temperature, giving support to the proposed
generalization scheme.

\end{abstract}

\maketitle

The study of the dynamics in non-equilibrium systems is an
intriguing and fascinating area of modern physics. There is an
obvious interest in the off-equilibrium regime as condensed matter
is often found far from ideal equilibrium condition. In these
cases where ``things keep happening on all time scales''
\cite{Fey} some of the most fundamental symmetries of equilibrium
statistical mechanics are broken. A particular class of systems 
where the out-of-equilibrium status occurs naturally 
are the so called glass forming systems. These
systems include, for example, disordered spin systems close to the
spin-glass transition, supercooled molecular liquids,
and jamming colloidal solutions. In general the dynamics of these
systems is extremely sensitive to change of external
parameters so that, when they are a little cooled or densified,
the time required to re-establish equilibrium grows enormously
and an off-equilibrium regime is entered. Here the average
quantities becomes time dependent and the correlation and response
functions depend on two times, in this regime the system is
known to perform \emph{aging}).

Correlation functions and response functions are the basic
quantities through which one probes the dynamics of a system in
many-body statistical physics. These functions are closely
related in equilibrium by the \emph{fluctuation-dissipation
theorem} (FDT). The FDT \cite{Hansen} establishes a relationship
between the correlation function $C_{AB}(t)=\langle A(0)B(t)
\rangle$ of the spontaneous equilibrium fluctuations of the dynamic variables
$A$ and $B$ and the response function $\chi_{AB}(t)=\langle A(t)
\rangle / h$, describing the change in the average value of $A$
due to an infinitesimal external field $h$ that is coupled to the
variable $B$ in the perturbation Hamiltonian:

\begin{equation} \label{eq:FDT}
\chi_{AB}(t)=\beta \left [ C_{AB}(0)-C_{AB}(t) \right ].
\end{equation}

\noindent Here $\beta=1/k_B T$, where $T$ is the kinetic temperature of
the system and $A$ and $B$ are variables that have zero-mean in the
unperturbed case. In this formulation of the FDT the field $h$
introduces an energy contribution $\delta H= -h B$ in the system
Hamiltonian and it is switched on instantaneously at $t=0$ and
kept constant for $t>0$ as an Heaviside step-function. In the
following we will refer to auto-correlation function ($B=A$) and
 drop the label $AA$ in $C_{AA}(t)$ and $\chi_{AA}(t)$.

Out of equilibrium the system is \emph{non-stationary} and
\emph{time-translational invariance} is lost. The correlation and
the response become two-times quantities depending also on the
\emph{aging time} $t_w$: $C=C(t_w,t_w+t)$, $\chi=\chi(t_w,t_w+t)$
and the FDT (\ref{eq:FDT}) is not supposed to hold. The
importance of extending the theorem to the non-equilibrium regime
has led to propose the \emph{generalized
fluctuation-dissipation relation} (GFDR) \cite{Crisanti,
Vulpiani, Cugliandolo}.

The generalization of the FDT proposed by Kurchan and Cugliandolo
\cite{Kurchan, Kurchan2} in the early '90 can be
expressed as follows:

\begin{equation} \label{eq:FDR}
\chi(t_w,t_w+t)=\beta \int_{t_w+t}^{t_w} ds
X(t_w,s)\partial_{s}C(t_w,s)
\end{equation}

\noindent where for short times $t$ (i.e. $t/t_w \ll 1$) the
previous equation reduces to the FDT (Eq.~\ref{eq:FDT}) and the
system is said to be in a \emph{quasi-equilibrium state}; for
intermediate $t$ ($t/t_w \simeq 1$) the violation function
$X(t_w,t)$ quantifies the deviation of the GFDR from the FDT;
finally, when $t/t_w \gg 1$, the function $X$ depends on $t_w$ and
$s$ only through the function $C$. The violation function can also
be interpreted \cite{theoryofTeff, detailedTeff} in terms of an
\emph{effective temperature} $T_{eff}=T/X$. The latter has
complicated behavior for intermediate $t$ values, but in the
long-$t$ regime it assumes a value that depends only on the waiting
time $t_w$ and one can think of it as "waiting-time dependent
effective temperature".

The GFDR in out-of-equilibrium systems has been studied
theoretically and through computer simulations in spin-glasses
\cite{SpinGlasses, Mezard} and in models for glassy dynamics\cite{Gregor}. 
Later, off-equilibrium molecular
dynamics simulations have given the possibility to perform the
same analysis on structural glasses \cite{Parisi1, Sellitto, Kob, Di
Leonardo}. Recently the generalized relation has also been
investigated in more exotic systems, like in the simulations for a
glassy protein \cite{bio}, in self-assembling processes of viral
capsids formation and of sticky disks crystallization
\cite{virdis} and in active systems composed of self-propelled
particles \cite{active}.

On the other hand few experiment have attempted to
study the GFDR\cite{Ocio,Grigera,
Bellon, Jabbari, Greinert, Ciliberto2, CilibertoNew}. Furthermore the results
reported until now for structural glasses remain controversial\cite{Grigera,
Bellon, Jabbari, Greinert, Ciliberto2}. The experiments in
structural glasses are intrinsically difficult because it is
necessary to simultaneously measure a correlation function of a
given variable and the associated response function, and all this
in a systems that is {\it instantaneously} brought out-of
equilibrium. These difficulties are relaxed in the case of
colloidal glasses (or gels, or jams), because the associated
timescales are much longer and fall easily in the experimentally
accessible window.

In this letter we report an experimental investigation of the
generalized FDR in a colloidal systems composed of a water
suspension of clay platelets (Laponite), which is {\it off
equilibrium} when it ages
towards the final arrested state (gel or glass, depending on the
clay concentration \cite{BR}). We study the reorientational
dynamics of the asymmetric clay platelets, looking at the
orientational {\it correlation functions} via depolarized dynamic
light scattering and at the corresponding {\it response function}
via the electric field induced birefringence. Measuring both
$C=C(t_w,t_w+t)$ and $\chi=\chi(t_w,t_w+t)$ at different waiting time
during the (days long) aging process we find that FDR is strongly
violated in the early stage of the aging process and is gradually
recovered as the aging time increases. Moreover, from the
parametric $C-\chi$ plot (the so called "FDT plot") we determine
the effective temperature and follow its evolution from the high
values ($T_{eff}/T \approx 5$) pertaining to young systems towards
the equilibrium ($T_{eff}/T=1$) attained at long waiting time. Our
findings confirm the generalization of the FDT to
off-equilibrium systems proposed by Cugliandolo and Kurchan 15
year ago.

{\it The off-equilibrium sample}. A solution, prepared by stirring
the Laponite powder with water, evolves toward an arrested state
on a timescale that span the ranges hours to months when simply kept
at room temperature and pressure. Even low concentration aqueous
solutions of this colloid, as the one used in our experiment
($\sim 1$ \% Laponite weight fraction), show strong aging 
of its light-scattering correlation function
\cite{Laura}. Due to its long (with respect to the experimental
timescale and to the decorrelation time) aging process of the
systems, one can approximate the different measure of fluctuations
and response as obtained in a sequence of steady out of
equilibrium states. Furthermore, the anisotropic shape of the clay disc makes it
possible to study its reorientation dynamics through the response
and the correlation function. Laponite particles dissolved
in water have the form of flat cylinders with a diameter of 25 nm
and an thickness of about 1 nm. Laponite colloidal particles are
good scatterers of visible light and this allow us to rapidly
measure the autocorrelation function of the scattered field.

{\it The correlation function}. In a dynamic light scattering
experiment one measures the correlation function of the optical
field scattered by the sample. The scattered field can be directly
related to the translational and rotational motion of the
anisotropic colloids suspended in the solvent~\cite{Pecora}. The
colloid's rotations are related to the second rank tensor of the
optical susceptibility. Specifically, in the VH (depolarized)
scattering geometry, one measures the autocorrelation function of
a variable that depends on the platelet's orientation:
\begin{equation}\label{eq:var}
A(t)=\sum_i P_{2}(\cos(\theta_i(t)))
\end{equation}

\noindent where $P_2(x)=(3x^2-1)/2$ is the 2nd order Legendre
polynomial, $\theta_i$ is the angle formed by the symmetry axes of
the $i$-th particle with the polarization vector of the incident
field and the sum is extended over the particles contained in the
scattering volume~\cite{Pecora}. This results holds exactly only
if the time scale of the rotational dynamics is much faster than
the translational one. This assumption was confirmed
by comparing the VV (polarized) and VH (depolarized) photon
correlation (PCS) at different waiting times and clay
concentration \cite{Supp}. The autocorrelation function of $A$ was
 measured using the VH geometry via
PCS. Several autocorrelation functions have been measured during
the aging process of the sample with a time resolution (1 $\mu$s)
dictated by the time-structure of the detector (photomultiplier)
response.

{\it The response function}. If one applies an external field that
tend to align the particle the system -due to the anisotropic
platelet's polarizability- becomes birefringent \cite{Boyd,Hecht}.
If the aligning field is a DC (or low frequency) electric field
(Kerr effect) the degree of rotation of a linearly polarized laser
beam is proportional to the square of the electric field via a
coefficient that is proportional to $A$ (Eq.~\ref{eq:var}).
Therefore, the (time dependent) Kerr response to the switch-on of
an electric field is proportional to the desired response function
(i.e. the response conjugated to the correlation function measured
in depolarized PCS). For selected values of the waiting times, the time resolved
response functions and the corresponding correlation functions,
was measured during the aging process of the Laponite
solution. The length of the electric pulses produced sets
the dynamic window of our experiment to about 1 ms.

 \begin{figure}
\begin{center}
\includegraphics[width=9.4cm]{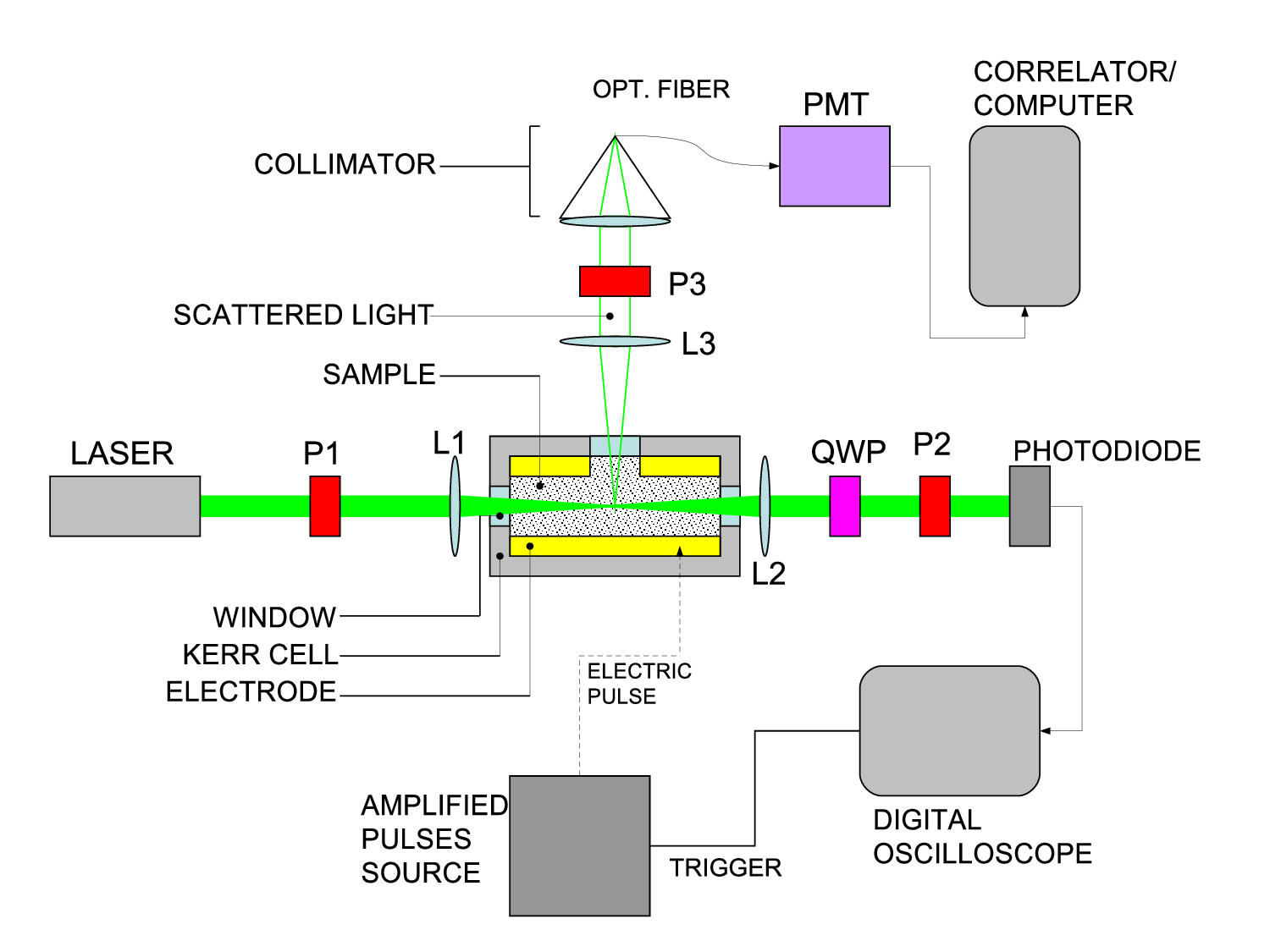}
\end{center}
\caption{\footnotesize{Sketch of the experimental set-up (see Ref.\cite{Supp} for more details). 
The laser's radiation ($\lambda=532$ nm) is polarized by the polarizer P1 and focused by the lens L1 at the centre of the cell containing the sample. The scattered light is collected by the lens L3 and selected by the polarizer P3 (orthogonal to P1). A photo-multiplier tube (PMT) detect the scatterd photons. When no electric pulse is applied to the cell the output of the PMT is acquired by a computer equipped with a custom digital correlator, this measures and stores the correlation function. The Kerr-cell containing the sample is provided by two electrodes connected to a source of amplified electric pulses. The forward-scatterd light (rotated by the electrically stimulated sample) is collected by the lens L2 and passes through a quarter-wave plate (QWP) and the polarizer P2 (orthogonal to P1). The transmitted light is detected by a photodiode connected to a digital oscilloscope. This is triggered to the source of electric pulse measuring and storing the Kerr-response function.}} \label{fig:s}
\end{figure}

Note that the relaxation time $\tau$ of these functions is always
much smaller that the typical waiting time ($\tau \ll t_w$). This
means that the time-resolved correlation and response are well-defined 
quantities although the system is aging. In addition, if
any FDT-violation can be detected, we expect to find
that in on a timescale comparable to the relaxation time
($t/\tau \geq 1$).

\begin{figure}
\begin{center}
\includegraphics[width=10.3cm]{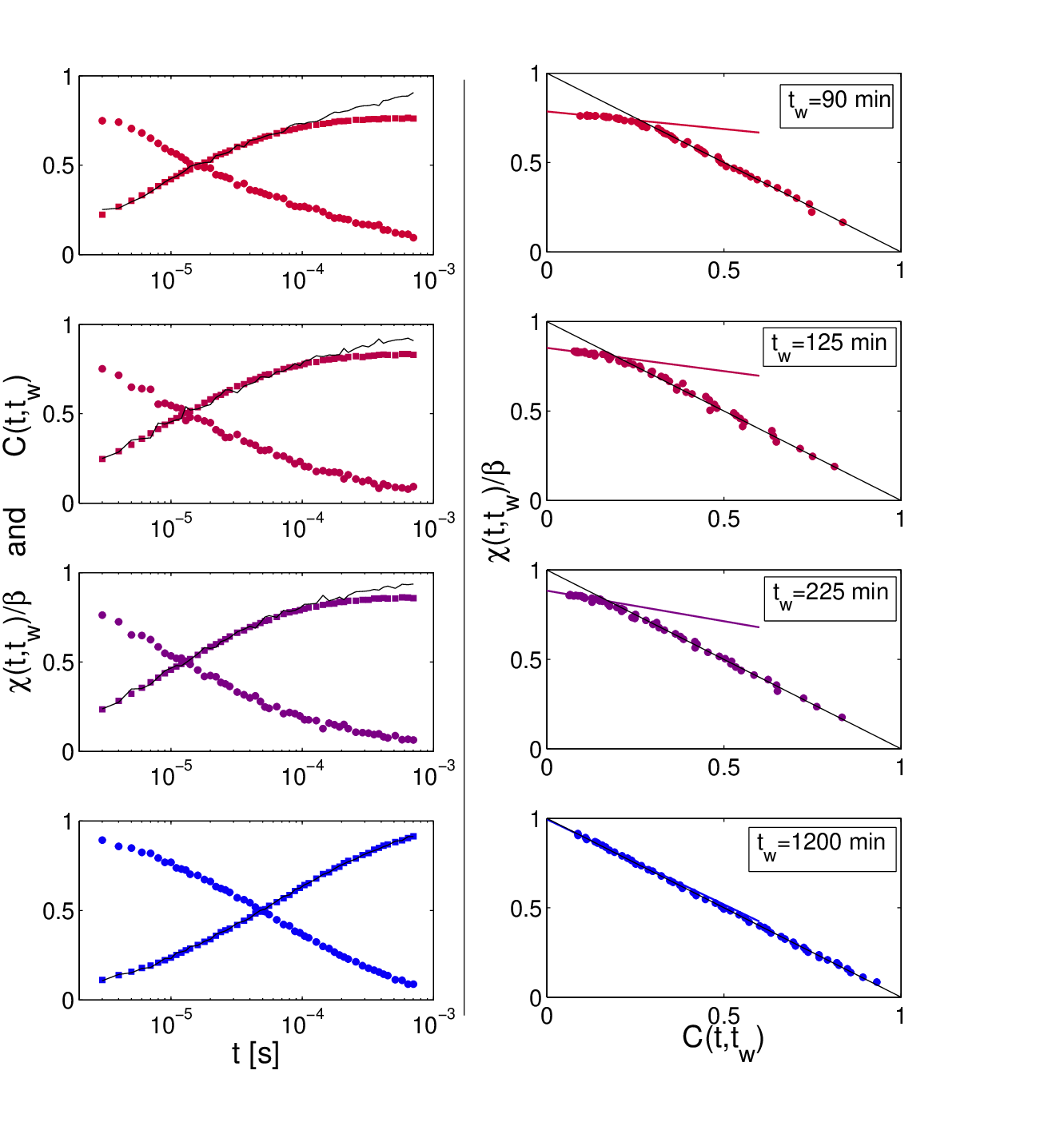}
\end{center}
\caption{\footnotesize{\textbf{(Left)} Normalized time correlation
function ($\circ$) and response function ($\Box$) measured at
different aging times (from top to bottom: $t_w=90$, 125, 225 and
1200 min), the solid line represents $1-C$. Note that $1-C$
deviates from $\chi/\beta$ when the FDT is violated. The
importance of this deviation reduces as $t_w$ increases.
\textbf{(Right)} Response function vs correlation function
measured at different aging times ($\circ$), the black line
represents the expectation of the FDT while the color lines
represent the linear fits to the points in the off-equilibrium
section of the FDT plot. It can be appreciated how these points
corresponding to long times approach gradually the FDT as the
aging time increases, for the longest waiting time the fitting
line overlaps almost perfectly with the prediction of the FDT (see
Fig.~\ref{fig:A} for a comparison of the FDT plots at different
aging times).}} \label{fig:FDT}
\end{figure}

\begin{figure}
\begin{center}
\includegraphics[width=10.1cm]{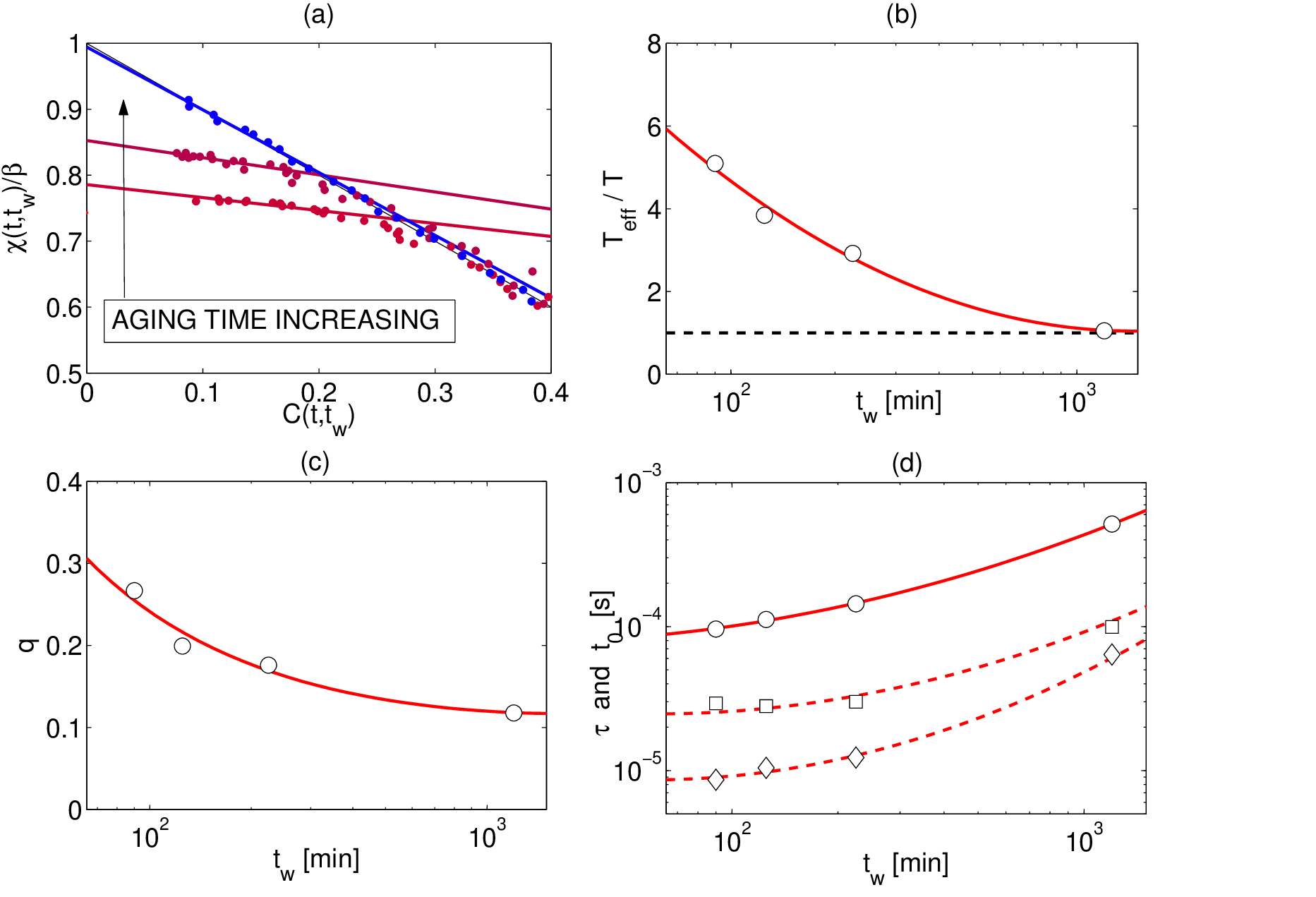}
\end{center}
\caption{\footnotesize{ \textbf{(a)} The interesting region of the
FDT plots for different aging times: $t_w=90$, 125 and 1200 min
(cfr. Fig.~\ref{fig:FDT}, right panel). \textbf{(b)} The inverse
slope of the long times points of the FDT plot (see Fig~.
\ref{fig:FDT}) as a function of the aging time, the red line is a
guide for the eye. This parameter can be interpreted as an
effective temperature and it is found to reduce to the bath
temperature as the aging time increases. \textbf{(c)} Waiting time
dependence of the Edwards-Anderson parameter, the characteristic
value of the correlation function at which the FDT breaks-down.
\textbf{(d)} Evolution with $t_w$ of the characteristic time at
which the FDT is violated ($\circ$), together with the relaxation
time of the correlation function ($\Box$) and of the response
($\diamondsuit$). The characteristic time for the FDT violation
$t_o$ increases as $t_w$ grows followed by the two relaxation
times.} } \label{fig:A}
\end{figure}

Examples of the measured quantities are shown in the left panel of
Fig.~\ref{fig:FDT}. The correlation function and the response
function are reported as functions of $t$ for different aging
times $t_w$. For short $t$ the FDT holds while we can see a clear
deviation from the FD relation for long $t$ where $\chi/\beta$
does not overlap with $1-C$ ($C(t)$ is normalized to $C(0)=1$).
When $\chi(t_w,t_w+t)/\beta$ is parametrically plotted against
$C(t_w,t_w+t)$ using $t$ as parameter (FDT plot) the departure from
the $1-C$ line becomes evident (see the right panel of
Fig.~\ref{fig:FDT}). The deviation from the behavior expected from
the FDT reduces its importance as $t_w$ grows, and the time where $T\chi$ and $C$ detaches from each other 
moves to longer
$t$  (see also Fig.~\ref{fig:A}(a), where the interested region of the FDT
plot has been expanded). In order to quantify this deviation, we
have performed a straight line fit to the longer time points in
the FDT plot. The slopes ($m$) of these lines are a measure of the
effective temperature: $1/m=T_{eff}/T$.

The $t_w$ dependence of $T_{eff}$ is reported in
Fig.~\ref{fig:A}(b): $T_{eff}$ decreases as $t_w$ increases. The
linear fit to the long $t$ region of the FDT plot also defines a
characteristic value of the correlation $C$ where the FDR
breaks-down, the so called Edwards-Anderson value $q$; this
quantity is reported as a function of $t_w$ in
Fig.~\ref{fig:A}(c). Finally, the quantity $q$, via $C(t_o)=q$,
identifies a characteristic time $t_o$ that mark the "starting
time" of the violation. $t_o$ is found to move to higher values as
the aging time increases (Fig.~\ref{fig:A}(d)). It is interesting
to compare $t_o$ to the relaxation times $\tau$ of the
response and the correlation. We find $\tau$ fitting the
correlation and the response with stretched exponential of the
form $\exp[-(t/\tau)^\beta]$ and $(1-a
\,\exp[-(t/\tau)^\beta])$, respectively. The response ages
faster than the correlation almost reaching the same relaxation
time for the longest $t_w$.

It is important to emphasize that in all models investigated so
far, for studying the generalization of the FDR, the relaxation time grows
roughly as the waiting time: $\tau \sim t_w$. The aging process
that we study experimentally here does not obey this simple
scaling, the relaxation time being  several orders of magnitude
shorter than the typical values of $t_w$. Nevertheless the
findings that we report in this work indicate that the generalized
form of theorem applies if $t/t_w$ is replaced by $t/\tau$ in
marking the transition of the different interesting regions.


In conclusion, by measuring the autocorrelation function of a
given variable and the response function of the same quantity in
an off-equilibrium (aging) colloidal suspension in the route to
the arrested state, we have tested the validity of the generalized
fluctuation-dissipation relation. The prediction of the GFDR apply
to the present experiment, and on the probed time-scale we
observe that the deviation from the satndard FDT reduces
gradually as the arrested phase is approached. The characteristic
time at which the violation is seen is always slightly above the
relaxation time of the measured response and correlation function.

\end{document}